\documentclass[
reprint,
floats,
showkeys,
showpacs,
superscriptaddress,
reprint,
amsfonts,
amssymb,
aps,
pra,
]{revtex4-1}

\newcommand{\Th}[2]{$^{#1}\mathrm{Th}^{#2}$}
\newcommand{\cm}{$\mathrm{cm}^{-1}$~}
\newcommand{\cmx}{$\mathrm{cm}^{-1}$}
\newcommand{\J}[1]{$J = #1$~}
\newcommand{\Jx}[1]{$J = #1$}
\newcommand{\lvl}[2]{$#1_{#2}~\mathrm{cm}^{-1}$~}
\newcommand{\lvlx}[2]{$#1_{#2}~\mathrm{cm}^{-1}$}

\usepackage[T1]{fontenc}
\usepackage{natbib}
\usepackage{graphicx}
\usepackage{dcolumn}
\usepackage{bm}
\usepackage{geometry}
\usepackage{color}
\usepackage{siunitx}
\usepackage{longtable}
\usepackage{bigstrut}
\usepackage{hyperref}

\DeclareSIUnit{\keV}{\kilo\eV}
\DeclareSIUnit{\MeV}{\mega\eV}

\begin{document}

\title{Electronic level structure of \Th{}{+} in the range of the \Th{229m}{} isomer energy}

\author{D.-M. Meier}\affiliation{Physikalisch-Technische Bundesanstalt, 38116 Braunschweig, Germany}
\author{J. Thielking}\affiliation{Physikalisch-Technische Bundesanstalt, 38116 Braunschweig, Germany}
\author{P. G\l{}owacki}\altaffiliation{Present address: Pozna\'{n} University of Technology, Pozna\'{n}, Poland}
\author{M. V. Okhapkin}\affiliation{Physikalisch-Technische Bundesanstalt, 38116 Braunschweig, Germany}
\author{R. A. M\"uller}\affiliation{Physikalisch-Technische Bundesanstalt, 38116 Braunschweig, Germany}
\author{A. Surzhykov}\affiliation{Physikalisch-Technische Bundesanstalt, 38116 Braunschweig, Germany}\affiliation{Technische Universit\"at Braunschweig, 38106 Braunschweig, Germany}
\author{E. Peik}\email[Corresponding author e-mail address: ]{ekkehard.peik@ptb.de}\affiliation{Physikalisch-Technische Bundesanstalt, 38116 Braunschweig, Germany}

\date{\today}

\begin{abstract}
Using resonant two-step laser excitation of trapped \Th{232}{+} ions, we observe 166 previously unknown energy levels of even parity within the energy range from \SI{7.8}{} to \SI{9.8}{eV} and angular momenta from \J{1/2} to $7/2$. We also classify the high-lying levels observed in our earlier experiments by the total angular momentum and perform multiconfiguration Dirac-Fock (MCDF) calculations to compare their results with the observed level density.
The observed levels can be relevant for the excitation or decay of the \Th{229m}{} isomeric nuclear state which lies in this energy range. The high density of electronic levels promises a strongly enhanced electronic bridge excitation of the isomer in \Th{229}{+}.
\end{abstract}



\maketitle


\section{Introduction}\label{sec:intro}

The atomic structure of thorium, the third-heaviest naturally occurring element, has been a subject of study because of applications in various fields. With four valence electrons in neutral \Th{}{} and strongly mixed 5f, 6d, 7s and 7p configurations the atomic spectrum is dense and complex~\cite{Actinides2006}. The high density of lines makes thorium discharge lamps suitable for the calibration of spectrographs over wide spectral ranges, for example in astronomical observatories~\cite{Lovis2007, Sansonetti2014}. In a recent analysis of the spectra of \Th{}{}, \Th{}{+} and \Th{}{2+}, Ritz wavelengths of about 20000 lines are given together with 787 energy levels of \Th{}{} and 516 of \Th{}{+}~\cite{Sansonetti2014}.

In our earlier work~\cite{Herrera2013} on \Th{}{+} we have seen the onset of repulsion in the energy spacing between levels of identical parity and angular momentum, as it is expected for strong configuration mixing, leading to the appearance of chaotic quantum dynamics~\cite{Rosenzweig1960}.

Work on excited states in \Th{}{+} is motivated by the search for laser excitation of the \Th{229m}{} nuclear isomer at \SI{7.8}{eV} excitation energy. This energy has been inferred from $\gamma$-spectrocopy~\cite{Beck2007, Beck2009} with an uncertainty $\sigma =0.5$~eV, that is large on the scale of optical spectroscopy.  No direct optical spectroscopy of the nuclear transition has been possible so far. Since the electronic level density in this energy range is high, it can be expected that laser excitation of the isomer and radiative decay are dominated by so called electronic bridge processes, where the electron shell enhances the coupling of the nucleus to the radiation field~\cite{Tkalya1996, Karpeshin1999, Porsev2010, Peik2015}, determined by the detuning between nuclear and electronic resonance. Therefore it is likely to find strong enhancement in a system with high electronic level density.
A tentative experimental indication of this mechanism could be conjectured from the observation of a strongly reduced lifetime of the isomer in \Th{}{+} in comparison to \Th{}{2+} and \Th{}{3+}~\cite{Seiferle2017,Karpeshin2018}.

\begin{figure}[h!tb]
\centering
\includegraphics[width=0.4\textwidth]{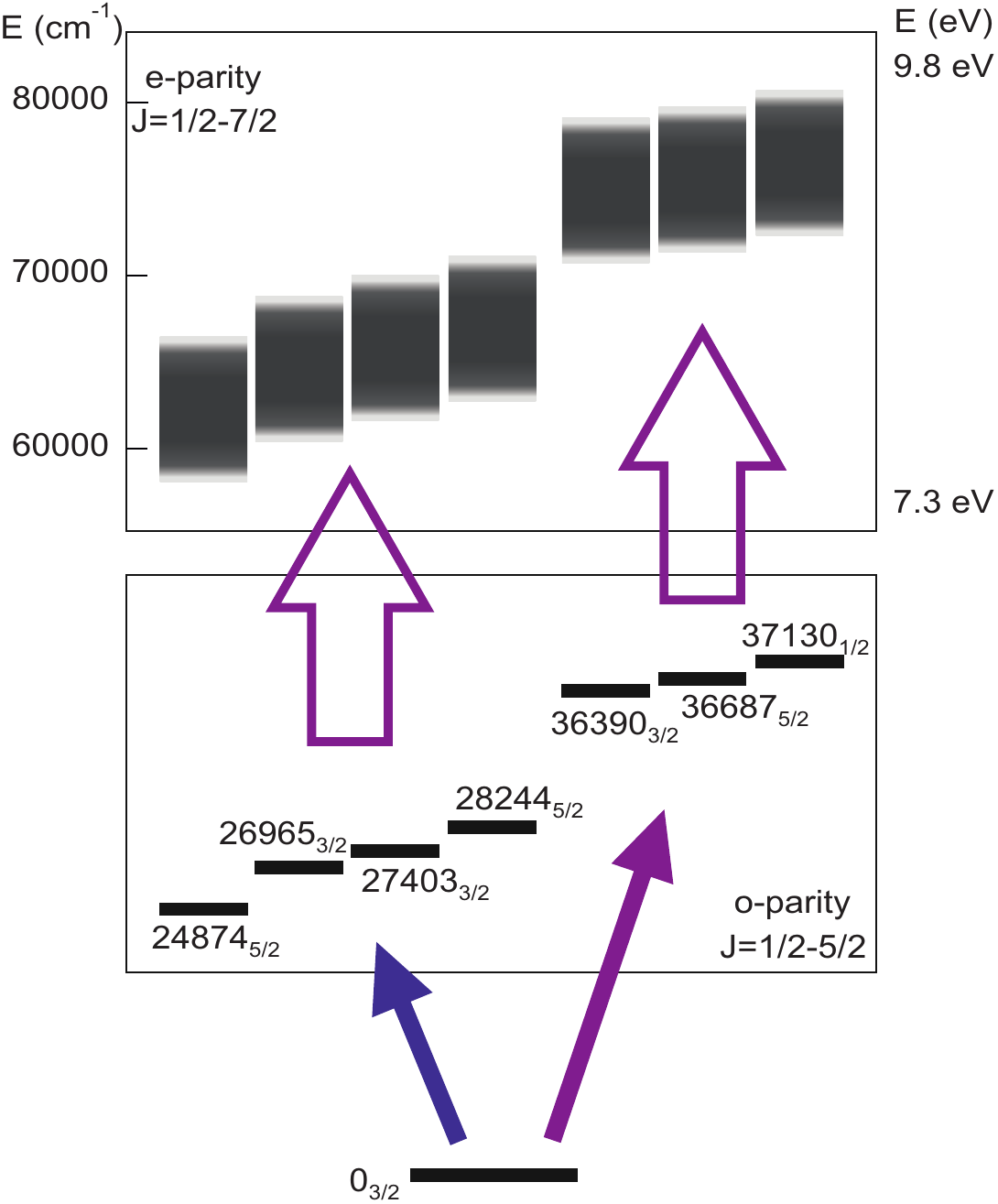}
\caption{
Two-step laser excitation scheme in \Th{}{+}, starting from the ground state through different odd-parity intermediate levels (middle panel) to  high-lying levels within the indicated energy ranges (upper panel). Levels are labelled by their energy in \cm and their total angular momentum $J$.}
\label{fig:level}
\end{figure}

In a previous publication~\cite{Herrera2013} 44 unknown energy levels of \Th{}{+} were reported within the energy range from \SI{7.3}{} to \SI{8.3}{eV}, obtained from two-step laser excitation of trapped \Th{}{+} ions. Extending this study, we present here 166 additional levels of even parity up to \SI{9.8}{eV} energy and classify these and the levels reported earlier according to their total angular momentum. In order to excite different configurations and different values of the angular momentum, 15 different intermediate levels are used as first excitation steps from the ground state.
Figure~\ref{fig:level} gives a schematic overview of the excitation schemes and Table~\ref{tab:config} lists the leading electronic configurations of the intermediate levels used in our experiments.

\begin{table}[htbp]
  \centering
  \caption{Intermediate levels of the two-step excitation, sorted by angular momentum $J$. Energies and leading configurations are taken from Ref.~\cite{Zalubas1974}, with the exception of the corrected classification of the 25027\cm state described in Ref.~\cite{Okhapkin2015}. Labels in the first column are introduced for reference in Tab.~\ref{tab:newlevels} in the appendix.}

    \begin{tabular}{|cccc|}
    \hline
    Ref. & Level [cm$^{-1}$] & J & Configuration \bigstrut[t]\\
	\hline
    a     & 25027.040 & 1/2   & $7s7p+6d7s7p$ \\
    b     & 26626.478 & 1/2   & $6d7s(^3\mathrm{D})7p\;^4\mathrm{D}^\mathrm{o} + (^3\mathrm{D})\;^2\mathrm{P}^\mathrm{o}$ \\
    c     & 35198.990 & 1/2   & $6d^2(^3P)7p\; ^4\mathrm{D}^\mathrm{o} + 6d7s(^3\mathrm{D})7p\; ^4\mathrm{P}^\mathrm{o}$ \\
    d     & 37130.340 & 1/2   & $5f(^1\mathrm{G}^\mathrm{o})6d^2\; ^2\mathrm{P}^\mathrm{o} + 6d^2(^3\mathrm{P})7p\;^4\mathrm{D}^\mathrm{o}$ \\
    e     & 37716.322 & 1/2   & $6d^2(^1\mathrm{D})7p\;^2\mathrm{P}^\mathrm{o} + (^3\mathrm{F})\;^4\mathrm{D}^\mathrm{o}$ \\
    f     & 26965.202 & 3/2   & $5f(^3\mathrm{F}^\mathrm{o})6d^2\;^4\mathrm{S}^\mathrm{o} +  (^3\mathrm{F}^\mathrm{o})\; ^2\mathrm{P}^\mathrm{o}$ \\
    g     & 27403.180 & 3/2   & $5f(^1\mathrm{D}^\mathrm{o})6d^2\;^2\mathrm{D}^\mathrm{o} + (^3\mathrm{P}^\mathrm{o})\;^4\mathrm{D}^\mathrm{o}$ \\
    h     & 36390.527 & 3/2   & $6d^2(^3\mathrm{F})7p\;^4\mathrm{F}^\mathrm{o} + 5f(^1\mathrm{G}^\mathrm{o})6d^2\; ^2\mathrm{P}^\mathrm{o}$ \\
    i     & 36581.568 & 3/2   & $6d^2(^3\mathrm{F})7p\;^4\mathrm{F}^\mathrm{o} + 5f(^1\mathrm{G}^\mathrm{o})6d^2\; ^2\mathrm{D}^\mathrm{o}$ \\
    j     & 24873.981 & 5/2   & $6d7s(^3\mathrm{D})7p\;^4\mathrm{F}^\mathrm{o} + 5f(^3\mathrm{P}^\mathrm{o})6d^2\; ^4\mathrm{F}^\mathrm{o}$ \\
    k     & 26424.480 & 5/2   & $5f(^3\mathrm{P}^\mathrm{o})6d^2\;^4\mathrm{F}^\mathrm{o} + (^3\mathrm{P}^\mathrm{o})\;^4\mathrm{D}^\mathrm{o}$ \\
    l     & 28243.812 & 5/2   & $6d7s(^3\mathrm{D})7p\;^4\mathrm{F}^\mathrm{o} + 6d^2(^3\mathrm{F})7p\;^4\mathrm{G}^\mathrm{o}$ \\
    m     & 35156.916 & 5/2   & $5f(^1\mathrm{G}^\mathrm{o})6d^2\; ^2\mathrm{D}^\mathrm{o} + (^1\mathrm{G}^\mathrm{o})\;^2\mathrm{F}^\mathrm{o}$ \\
    n     & 36687.992 & 5/2   & $6d^2(^3\mathrm{F})7p\;^4\mathrm{F}^\mathrm{o} +  6d7s(^3\mathrm{D})7p\;^4\mathrm{P}^\mathrm{o}$ \\
    o     & 37846.174 & 5/2   & $6d^2(^3\mathrm{P})7p\;^2\mathrm{D}^\mathrm{o} + (^3\mathrm{P})\; ^4\mathrm{D}^\mathrm{o}$ \bigstrut[b]\\
    \hline
	\hline
    \end{tabular}%
	\label{tab:config}%
\end{table}%

The range of excitation energy that is investigated here covers more than $2\sigma$ above the \Th{229m}{} isomer energy of \SI{7.8 \pm 0.5}{eV}, but is still well below the ionization potential of \Th{}{+} of 12.1(2) eV~\cite{Herrera2013}.
A further increase in level density by up to a factor ten has been predicted in approaching the ionization energy~\cite{Dzuba2010}.
Multi-configurational Dirac-Fock (MCDF) calculations are used to estimate the level density in the range of the excitation energy of the nuclear isomer.

\section{Experimental}\label{sec:experimental}

For the experiments we use a linear Paul trap with the capacity to store up to $10^6$ \Th{232}{+} ions, described in Ref.~\cite{Herrera2012, Herrera2012b}. The trap is loaded by ablating a metallic \Th{232}{} target using a Nd:YAG laser emitting \SI{5}{ns} pulses with an energy of \SI{\leq1}{mJ} at \SI{1064}{nm}. Argon buffer gas at \SI{0.1}{Pa} pressure is used to cool the trapped ions to room temperature and to depopulate metastable states by collisional quenching~\cite{Herrera2012, Herrera2012b}. We dissociate \Th{}{+} molecular compounds which are formed in the trap with impurities of the buffer gas (see~\cite{Herrera2013}) with the $4^{th}$ harmonic radiation of a Q-switched solid state laser at \SI{266}{nm} and a pulse energy of \SI{\approx10}{\mu J}.

\begin{figure}[h!tb]
\centering
\includegraphics[width=0.45\textwidth]{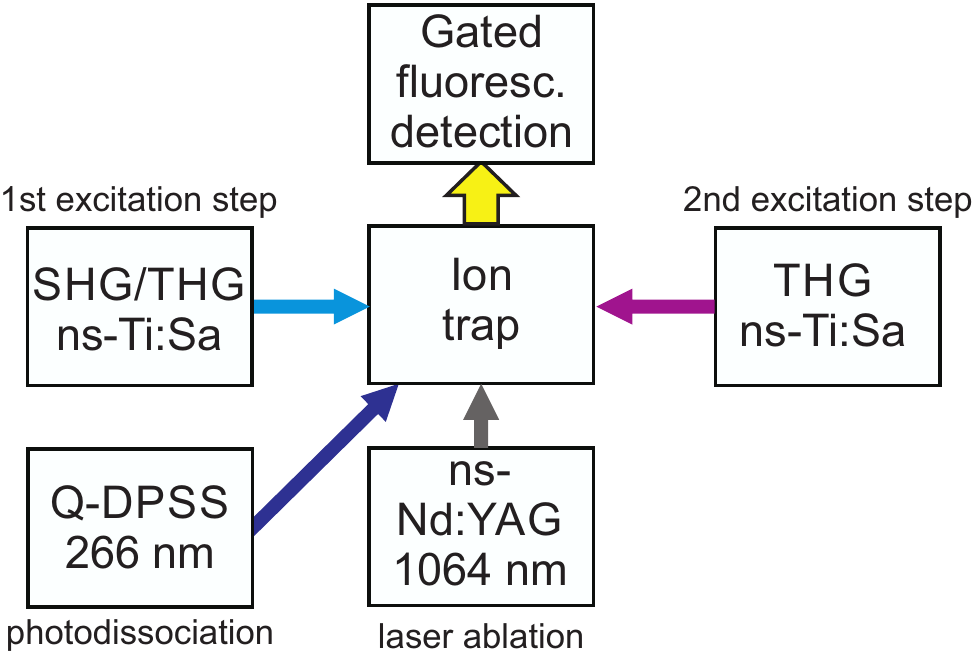}
\caption{Experimental setup for laser excitation of trapped \Th{}{+} ions. The first step excitation is provided by the second harmonic (SHG) or third harmonic (THG) radiation of a nanosecond Ti:Sa laser, locked to a Fizeau wavemeter. The second excitation step is provided by the THG of a second ns-Ti:Sa laser. \Th{}{+} ions are loaded via laser ablation (ns-Nd:YAG laser at \SI{1064}{nm}). Molecular compounds of \Th{}{+} are photodissociated by pulses from a Q-switched diode-pumped solid-state laser (Q-DPSS) at \SI{266}{nm}.}
\label{fig:opt_setup}
\end{figure}

The search of new levels is carried out by a two-step excitation: An odd-parity intermediate level (see Tab.~\ref{tab:config}) is excited from the ground state and the radiation frequency of a second laser is tuned over a wide frequency range to excite unknown even-parity high-lying levels from the intermediate level. A scheme of the experimental setup is shown in Fig.~\ref{fig:opt_setup}. The first excitation step is provided by the radiation of a second harmonic (SHG) or third harmonic generation (THG) of a pulsed Ti:Sa laser (TU model, Photonics Industries). For the second excitation step we use the radiation of a THG of a second pulsed Ti:Sa laser (Credo, Sirah Lasertechnik).

Both lasers emit synchronized pulses with a duration of \SI{\approx 20}{ns} and a repetition rate of \SI{1}{kHz}. The beam diameters of both lasers, which are matched at the position of the ions in the linear Paul trap, are \SI{\approx 1}{mm} and the peak powers used in the experiment are up to \SI{0.1}{kW}, which corresponds to intensities of \SI{10}{kW/cm^2}. The wavelengths of both lasers are measured by a Fizeau-wavemeter (HighFinesse/Angstrom WS-7). The first laser wavelength is stabilized to the center of the Doppler-broadened line of one of a set of selected intermediate levels by a computer-based locking scheme using the wavemeter readout. The second laser is tunable by controlling an intracavity diffraction grating and simultaneously adjusting the orientation of the frequency conversion crystals. The third harmonic radiation is tuned in the range from \SI{240}{} to \SI{293}{nm} at a tuning rate of \SI{0.001}{nm/s} in frequency steps of \SI{\approx 500}{MHz} (\SI{0.1}{pm}) or \SI{\approx 100}{MHz} for fine scans. In the search for new levels we determine the line center of each transition by repeated fine scanning over the line.

For the detection of the fluorescence decay of the high-lying electronic levels we use two photomultiplier tubes (PMT) for different spectral ranges. Photons are detected in the range from \SI{300}{} to \SI{650}{nm} with a PMT which is equipped with a long-pass edge interference filter to block laser stray light from the Ti:Sa laser pulses. A second PMT is used for the detection of VUV photons in the wavelength range from \SI{115}{} to \SI{230}{nm}.
Fast gated integrators are used to evaluate the PMT signals during a detection window of \SI{100}{ns}, starting after the excitation pulses. Counters are used to detect photons in the interval from \SI{1}{} to \SI{10}{\micro s} after the excitation pulses.

We try to excite each observed level via several intermediate levels with different values of the angular momentum (see Tab.~\ref{fig:level}) to confirm them as high-lying levels and determine their total angular momenta.

\section{Newly observed levels}\label{sec:two-photon}

Within the search scans we not only observe direct two-step excitation to high-lying levels, but also single-photon transitions originating from the ground state, thermally populated metastable levels (up to \lvlx{4490}{}) and levels populated by a decay of the intermediate level (N-scheme transition). Most of these transitions can be identified based on the existing energy level database~\cite{Zalubas1974}. In order to unambiguously identify new levels at an excitation energy corresponding to the sum of the two laser photon energies, we require to observe the excitation via at least two intermediate levels.

\begin{figure}[h!tb]
\centering
\includegraphics[width=0.5\textwidth]{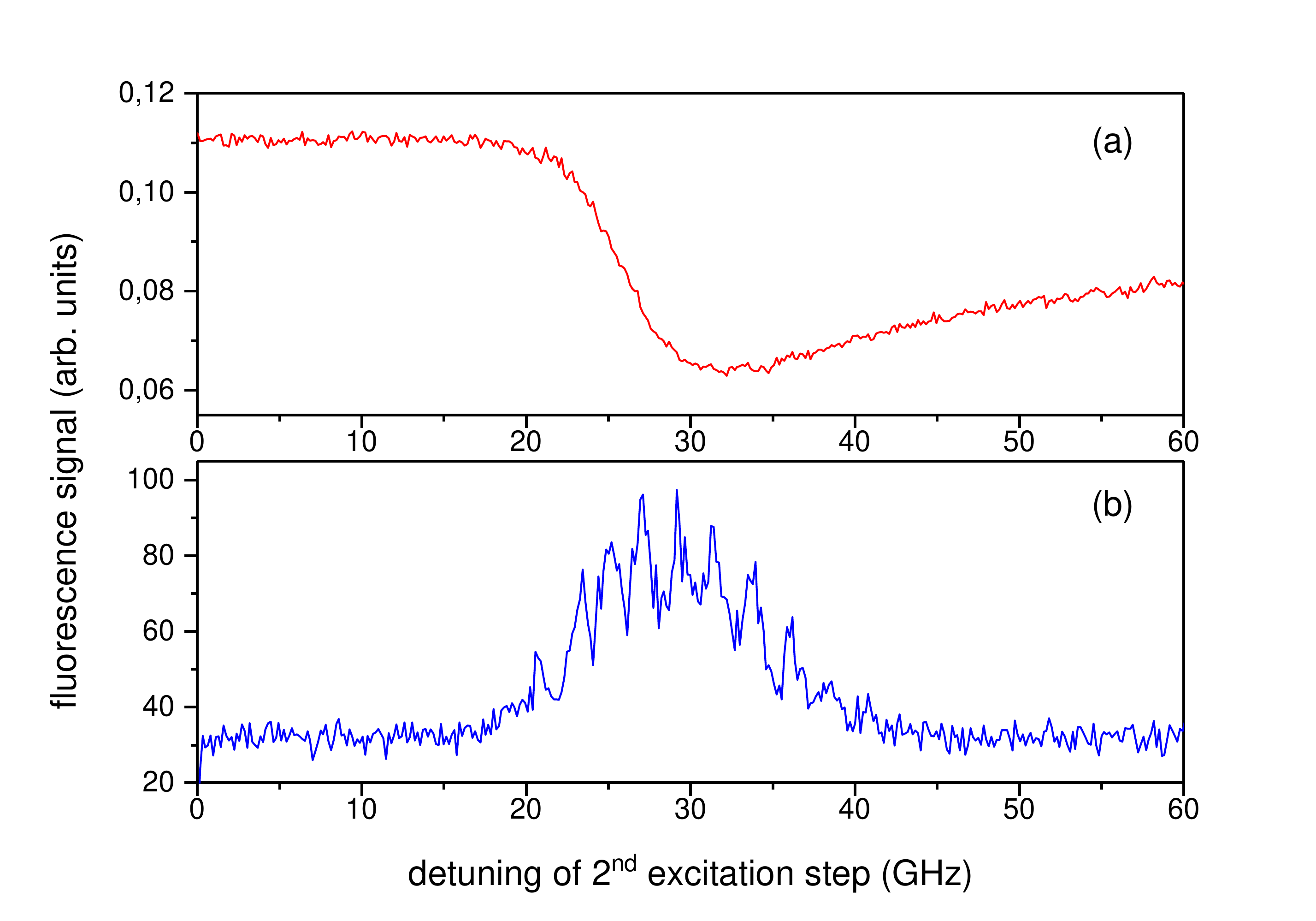}
\caption{Two-step excitation of the level \lvlx{74195.13}{} via the intermediate level \lvlx{35156.9}{5/2}. Graph (a) shows the decrease of the fluorescence signal of the intermediate level (wavelength range from \SI{325}{} to \SI{650}{nm}) due to three-photon ionization during the scan from negative to positive detunings. The slight increase of the signal after the excitation occurs due to redistribution of \Th{}{+} ions in the trap. Graph (b) shows the simultaneously recorded VUV fluorescence signal of the decay of the high-lying level. The lineshape is determined by the multimode spectrum of the Ti:Sa laser radiation. The start of the scan at \SI{0}{GHz} laser detuning corresponds to the wavenumber \lvlx{39039.15}{} of the $2^{nd}$ excitation step laser radiation.}
\label{fig:TwoStepEx}
\end{figure}

Excitation of a high-lying level is inferred by observing either fluorescence at a wavelength different from the excitation wavelength, depletion of fluorescence from the decay of the intermediate level or a loss of ions because of resonance-enhanced three-photon ionization, see Fig.~\ref{fig:TwoStepEx}. Due to the high pulse power of the Ti:Sa laser radiation there is a significant probability of absorption of a third photon when a high-lying level is excited. The absorption of this photon leads to ionization of the \Th{}{+} ions to \Th{}{2+} ions~\cite{Herrera2013}. We select the operating parameters of our ion trap so that \Th{}{2+} is not stably trapped.
Nevertheless, ionization alone cannot be a confirmation of the high-lying level excitation, because it can also occur as a resonant four-photon process in an N-scheme excitation through levels with an energy of about \lvlx{20000}{}, listed in Ref.~\cite{Zalubas1974}.

\begingroup
\LTcapwidth=\linewidth

\begin{longtable*}{|ccc|ccc|ccc|}

\caption{Newly found electronic levels in \Th{232}{+} in the energy range from \SI{7.8}{} to \SI{9.8}{eV}, total uncertainties $\sigma_{tot}$ of the energy determination and possible angular momenta. The angular momentum value \J{3/2} is assigned with certainty, otherwise the most likely values of $J$ are given (see text). Levels marked with * were found in the previous investigation~\cite{Herrera2013} with an uncertainty of 0.2 \cm and are classified in this investigation. We update the energy value of a previously found level, marked with **.}\\
\hline
    Energy [cm$^{-1}$] & $\sigma_{tot}$ [cm$^{-1}$] & J     & Energy [cm$^{-1}$] & $\sigma_{tot}$ [cm$^{-1}$] & J     & Energy [cm$^{-1}$] & $\sigma_{tot}$ [cm$^{-1}$] & J \bigstrut\\
\hline
\endfirsthead
\caption{Cont.: Newly found electronic levels in $^{232}\mathrm{Th}^+$.} \\
\hline
    Energy [cm$^{-1}$] & $\sigma_{tot}$ [cm$^{-1}$] & J     & Energy [cm$^{-1}$] & $\sigma_{tot}$ [cm$^{-1}$] & J     & Energy [cm$^{-1}$] & $\sigma_{tot}$ [cm$^{-1}$] & J \bigstrut\\
\hline
\endhead
\hline
\endfoot
\hline
\hline
\endlastfoot
	58875.5 & *     & 3/2 - 7/2 & 68564.19 & 0.18  & 3/2 or 5/2 & 74195.13 & 0.17  & 3/2 - 7/2 \bigstrut[t]\\
    59387.1 & *     & 3/2 - 7/2 & 68598.83 & 0.18  & 3/2 or 5/2 & 74201.68 & 0.16  & 1/2 or 3/2 \\
    59477.4 & *     & 3/2   & 68752.06 & 0.17  & 3/2   & 74255.11 & 0.17  & 3/2 or 5/2 \\
    59803.0 & *     & 3/2 - 7/2 & 68812.64 & 0.18  & 3/2 or 5/2 & 74328.90 & 0.17  & 3/2 or 5/2 \\
    60380.1 & *     & 3/2   & 68898.67 & 0.17  & 3/2 or 5/2 & 74396.40 & 0.16  & 3/2 or 5/2 \\
    60618.6 & *     & 3/2   & 68921.30 & 0.17  & 3/2 or 5/2 & 74461.06 & 0.17  & 3/2 - 7/2 \\
    60721.3 & *     & 3/2 - 7/2 & 69582.93 & 0.17  & 3/2   & 74503.38 & 0.18  & 3/2 or 5/2 \\
    61032.4 & *     & 3/2   & 70036.57 & 0.17  & 3/2   & 74554.39 & 0.16  & 3/2 \\
    61388.0 & *     & 3/2 - 7/2 & 70602.32 & 0.16  & 3/2 or 5/2 & 74642.51 & 0.17  & 3/2 or 5/2 \\
    61428.6 & *     & 3/2 or 5/2 & 70618.40 & 0.16  & 3/2 or 5/2 & 74646.94 & 0.16  & 1/2 or 3/2 \\
    61726.3 & *     & 3/2   & 70747.82 & 0.16  & 3/2 - 7/2 & 74781.76 & 0.17  & 3/2 or 5/2 \\
    61963.6 & *     & 3/2   & 70924.75 & 0.16  & 3/2 or 5/2 & 74823.59 & 0.16  & 3/2 or 5/2 \\
    62307.2 & *     & 3/2 or 5/2 & 71038.37 & 0.16  & 3/2   & 74881.76 & 0.17  & 3/2 \\
    62373.8 & *     & 3/2   & 71043.67 & 0.15  & 3/2 - 7/2 & 75009.01 & 0.18  & 3/2 or 5/2 \\
    62477.0 & *     & 3/2 or 5/2 & 71148.96 & 0.16  & 3/2 or 5/2 & 75058.06 & 0.17  & 3/2 - 7/2 \\
    62560.1 & *     & 3/2 or 5/2 & 71153.87 & 0.16  & 3/2   & 75108.36 & 0.17  & 3/2 or 5/2 \\
    62562.2 & *     & 3/2   & 71278.32 & 0.16  & 1/2 or 3/2 & 75129.42 & 0.17  & 1/2 - 5/2 \\
    62753.1 & *     & 3/2 or 5/2 & 71309.24 & 0.16  & 3/2 or 5/2 & 75172.74 & 0.19  & 3/2 or 5/2 \\
    62873.11 & 0.17  & 1/2 or 3/2 & 71345.59 & 0.16  & 1/2 or 3/2 & 75324.44 & 0.17  & 3/2 - 7/2 \\
    63257.5 & *     & 3/2 - 7/2 & 71457.82 & 0.16  & 3/2 - 7/2 & 75380.56 & 0.17  & 3/2 or 5/2 \\
    63268.90 & 0.17  & 1/2 or 3/2 & 71543.05 & 0.16  & 3/2   & 75434.83 & 0.17  & 3/2 - 7/2 \\
    63298.4 & *     & 3/2 - 7/2 & 71544.12 & 0.16  & 1/2 or 3/2 & 75553.70 & 0.17  & 1/2 or 3/2 \\
    63557.7 & *     & 3/2 - 7/2 & 71595.39 & 0.17  & 1/2 - 5/2 & 75568.80 & 0.17  & 3/2 or 5/2 \\
    63680.29 & 0.16  & 3/2   & 71648.63 & 0.16  & 1/2 - 5/2 & 75613.54 & 0.17  & 3/2 \\
    64107.51 & 0.17  & 3/2 or 5/2 & 71682.08 & 0.16  & 1/2 or 3/2 & 75690.72 & 0.17  & 1/2 or 3/2 \\
    64122.0 & *     & 3/2 - 7/2 & 71704.43 & 0.16  & 3/2 or 5/2 & 75783.69 & 0.17  & 3/2 or 5/2 \\
    64150.3 & *     & 3/2   & 71893.76 & 0.16  & 3/2   & 75840.96 & 0.16  & 3/2 or 5/2 \\
    64368.24 & 0.17  & 3/2 or 5/2 & 71980.22 & 0.17  & 3/2 or 5/2 & 75889.36 & 0.17  & 1/2 or 3/2 \\
    64442.11 & 0.16  & 1/2 or 3/2 & 71995.47 & 0.16  & 3/2 - 7/2 & 75950.56 & 0.17  & 3/2 - 7/2 \\
    64560.4 & *     & 3/2   & 72027.87 & 0.16  & 3/2   & 75966.37 & 0.16  & 3/2 \\
    64813.7 & *     & 3/2 - 7/2 & 72070.57 & 0.16  & 1/2 or 3/2 & 76122.62 & 0.16  & 1/2 or 3/2 \\
    64860.4 & *     & 3/2 or 5/2 & 72183.47 & 0.17  & 3/2 or 5/2 & 76157.73 & 0.16  & 1/2 or 3/2 \\
    64887.80 & 0.16  & 1/2 or 3/2 & 72195.77 & 0.17  & 3/2 or 5/2 & 76158.83 & 0.17  & 3/2 - 7/2 \\
    64920.1 & *     & 3/2 or 5/2 & 72353.67 & 0.17  & 3/2 - 7/2 & 76169.30 & 0.17  & 1/2 - 5/2 \\
    65037.7 & *     & 3/2   & 72395.99 & 0.18  & 3/2 or 5/2 & 76227.32 & 0.17  & 1/2 - 5/2 \\
    65144.4 & *     & 3/2 or 5/2 & 72403.45 & 0.16  & 3/2   & 76371.41 & 0.16  & 1/2 or 3/2 \\
    65191.1 & *     & 3/2 - 7/2 & 72437.78 & 0.16  & 1/2 or 3/2 & 76445.96 & 0.16  & 1/2 or 3/2 \\
    65730.4 & *     & 3/2   & 72443.68 & 0.16  & 3/2 or 5/2 & 76508.84 & 0.17  & 3/2 - 7/2 \\
    65738.54 & 0.18** & 3/2 - 7/2 & 72612.99 & 0.18  & 3/2 or 5/2 & 76521.57 & 0.17  & 3/2 or 5/2 \\
    65753.45 & 0.17  & 1/2 or 3/2 & 72624.83 & 0.16  & 3/2   & 76552.60 & 0.16  & 3/2 \\
    65799.6 & *     & 3/2   & 72644.97 & 0.16  & 3/2 - 7/2 & 76616.44 & 0.17  & 3/2 - 7/2 \\
    65910.0 & *     & 3/2 - 7/2 & 72711.86 & 0.16  & 3/2 - 7/2 & 76788.94 & 0.16  & 1/2 or 3/2 \\
    65946.9 & *     & 3/2 - 7/2 & 72748.31 & 0.16  & 3/2   & 76895.40 & 0.16  & 3/2 \\
    66052.0 & *     & 3/2 or 5/2 & 72756.38 & 0.17  & 3/2 - 7/2 & 76954.26 & 0.16  & 3/2 \\
    66141.2 & *     & 3/2 or 5/2 & 72849.84 & 0.16  & 3/2   & 76999.66 & 0.16  & 3/2 - 7/2 \\
    66324.52 & 0.17  & 1/2 or 3/2 & 72904.32 & 0.16  & 3/2 - 7/2 & 77069.42 & 0.17  & 3/2 or 5/2 \\
    66333.7 & *     & 3/2   & 72937.01 & 0.17  & 3/2 or 5/2 & 77154.88 & 0.17  & 1/2 - 5/2 \\
    66388.81 & 0.17  & 1/2 or 3/2 & 72967.30 & 0.16  & 3/2   & 77185.31 & 0.16  & 3/2 - 7/2 \\
    66429.64 & 0.17  & 3/2 or 5/2 & 73007.80 & 0.16  & 1/2 or 3/2 & 77208.86 & 0.17  & 3/2 or 5/2 \\
    66558.0 & *     & 3/2 - 7/2 & 73141.86 & 0.16  & 3/2   & 77278.24 & 0.16  & 1/2 or 3/2 \\
    66609.0 & *     & 3/2 or 5/2 & 73171.82 & 0.17  & 3/2 - 7/2 & 77295.28 & 0.17  & 3/2 or 5/2 \\
    66666.96 & 0.17  & 1/2 or 3/2 & 73225.52 & 0.17  & 3/2   & 77311.25 & 0.16  & 3/2 \\
    66702.9 & *     & 3/2   & 73245.52 & 0.17  & 3/2 or 5/2 & 77429.20 & 0.16  & 1/2 or 3/2 \\
    66831.1 & *     & 3/2   & 73314.63 & 0.16  & 3/2 - 7/2 & 77506.72 & 0.16  & 1/2 - 5/2 \\
    66855.6 & *     & 3/2 or 5/2 & 73349.33 & 0.18  & 1/2 or 3/2 & 77668.89 & 0.17  & 3/2 or 5/2 \\
    67066.2 & *     & 3/2 - 7/2 & 73427.59 & 0.16  & 3/2   & 77715.95 & 0.17  & 1/2 or 3/2 \\
    67154.05 & 0.16  & 3/2   & 73486.18 & 0.18  & 3/2 or 5/2 & 77914.52 & 0.16  & 1/2 or 3/2 \\
    67177.76 & 0.18  & 3/2 - 7/2 & 73506.20 & 0.16  & 1/2 or 3/2 & 77992.06 & 0.16  & 3/2 - 7/2 \\
    67378.61 & 0.19  & 3/2 - 7/2 & 73514.42 & 0.17  & 3/2 - 7/2 & 78004.65 & 0.16  & 3/2 \\
    67509.63 & 0.17  & 3/2   & 73571.40 & 0.17  & 3/2 or 5/2 & 78079.15 & 0.16  & 3/2 or 5/2 \\
    67577.71 & 0.20  & 3/2 or 5/2 & 73637.54 & 0.16  & 3/2 or 5/2 & 78106.22 & 0.16  & 1/2 or 3/2 \\
    67657.30 & 0.18  & 1/2 or 3/2 & 73717.33 & 0.16  & 3/2   & 78147.89 & 0.17  & 3/2 - 7/2 \\
    67737.62 & 0.18  & 3/2 or 5/2 & 73720.66 & 0.19  & 3/2 - 7/2 & 78311.97 & 0.17  & 1/2 or 3/2 \\
    67803.24 & 0.17  & 3/2 or 5/2 & 73856.76 & 0.16  & 3/2 or 5/2 & 78365.42 & 0.17  & 3/2 \\
    67843.31 & 0.17  & 3/2 or 5/2 & 73927.39 & 0.17  & 3/2 or 5/2 & 78486.07 & 0.17  & 1/2 or 3/2 \\
    67866.10 & 0.18  & 3/2 or 5/2 & 74015.17 & 0.17  & 3/2 - 7/2 & 78643.42 & 0.17  & 1/2 or 3/2 \\
    68033.33 & 0.21  & 3/2 - 7/2 & 74035.55 & 0.16  & 3/2   & 78671.36 & 0.16  & 1/2 or 3/2 \\
    68088.03 & 0.17  & 3/2   & 74041.31 & 0.16  & 3/2 - 7/2 & 78746.08 & 0.16  & 1/2 or 3/2 \\
    68278.65 & 0.20  & 3/2 - 7/2 & 74077.59 & 0.17  & 1/2 - 5/2 & 78780.27 & 0.17  & 1/2 or 3/2 \\
    68497.88 & 0.18  & 3/2 or 5/2 & 74159.76 & 0.17  & 3/2 or 5/2 & 79056.55 & 0.17  & 1/2 or 3/2 \bigstrut[b]
\label{tab:newlevelsshort}
\end{longtable*}
\endgroup

By applying the aforementioned criteria, two-photon excitations to 166 previously unknown high-lying energy levels in the investigated energy range from \SI{7.3}{} to \SI{9.8}{eV} are identified and listed in Tab.~\ref{tab:newlevelsshort}. In Tab.~\ref{tab:newlevels} given in the appendix a complete list with the successful excitation pathways via the different intermediate levels is provided. We achieve a total uncertainty for the energy determination of $\sigma_{tot} \sim 0.2$ \cmx, limited by the uncertainty of identifying the center of the Ti:Sa laser spectrum from the wavemeter readout. The energy of a high-lying level is derived as the mean value from the different excitation pathways, weighted with the statistical uncertainty of the individual measurements.
The measured center wavelengths of the transitions from the ground state to the intermediate levels mentioned in Tab.~\ref{tab:config} are in coincidence with the values given in Ref.~\cite{Zalubas1974}.

In comparison to the previous experiment~\cite{Herrera2013} we see an agreement of the measured energies of these levels, except for one level. The energy of the level marked with ** in Tab.~\ref{tab:newlevelsshort} is measured to be \lvlx{65738.54}{}, shifted by $\approx2\sigma$ from the previous value. We verified the existence of the level and update its energy to the new value, which is based on the excitations from three intermediate levels. A level mentioned at \lvl{66427.14}{} in Ref.~\cite{Sansonetti2014} is not observed in our experiment.

\section{Determination of the total angular momenta}

The observed signal strenghts at the applied second step laser intensities in the range of \SI{10}{kW/cm^2} are in the range expected for electric dipole transitions~\cite{Herrera2013}. Therefore, we assume that the observed transitions are electric dipole allowed. Selection rules then require this high-lying levels to have even parity and angular momenta from \J{1/2} to $7/2$, since we are using intermediate levels with \J{1/2} to $5/2$.

Analyzing the excitation through different intermediate levels allows for the determination of the total angular momentum of the newly found levels. In our experiment high-lying levels with \J{3/2} can be accessed via all intermediate levels. These states can therefore be unambiguously classified. Angular momenta \Jx{1/2}, $5/2$ and $7/2$ can only be identified by the absence of excitation through intermediate levels that would not fulfil the $\Delta J=0,\pm1$ selection rule.

In these cases, the absence of a single excitation path does not allow a sure assignment of the $J$ value, as we observe that some high-lying levels can be excited through only one out of two intermediate levels with the same $J$ (see, e.g. level \lvlx{67577.7}{}, in Tab.~\ref{tab:newlevels} in the appendix). This is because intermediate levels with the same total angular momenta can have different electronic configurations (see Tab.~\ref{tab:config}) and therefore the excitation probabilities of the high-lying levels can be substantially different. Hence, we require the absence of excitation through at least two intermediate levels with the same $J$ to consider the level not being excitable via this particular $J$. In all other cases we give the range of possible angular momenta.

Using the same excitation scheme we also determine the angular momenta of levels which could not be classified in Ref.~\cite{Herrera2013}.

\section{Level density calculations}

To reassure that we did not miss a relevant number of levels in the experiment we calculated the densities of even levels in \Th{}{+} up to \lvl{85000}{} using the multi-configurational Dirac-Fock (MCDF) method implemented in a modified version of the \textsc{Grasp2k} package \cite{Jonsson2013}. A detailed description of the methods involved can be found e.g. in Refs. \cite{Fischer1997, Grant2007}.
Thorium in its singly ionized state has three valence electrons above a closed radon core. The basis for our calculations has been set up with single and double excitations of the valence electrons from six reference configurations: $6d^27s$, $5f^27s$, $5f^26d$, $5f7s7p$, $5f6d7p$ and $6d^3$. Moreover we add to the basis the configurations produced by single excitations from the core opened up to xenon.

Singly charged thorium is a very complicated system with strong electron-electron correlations. Therefore we cannot expect to reproduce the energies of particular levels accurately, especially for high-lying levels. Instead we compare the number of levels per \lvl{5000}{} interval. Generally the size of the basis used in an MCDF calculation determines the number of energy levels that can be constructed, each approximation a solution of the many-electron Dirac equation. However, due to numerical uncertainties, not all of these energy levels have a physical meaning. Especially those that consist only of basis elements constructed from weakly bound single-electron orbitals are numerically unstable. Fortunately such energy levels are collected at energies close to the ionization threshold. This can be checked by observing the convergence of the level density at different energy ranges. While the level density converges and is stable for the energy range under consideration in the present work, an increasing number of non-physical energy levels can be observed at higher energies.

\begin{figure}[h!tb]
\centering
\includegraphics[width=0.50\textwidth]{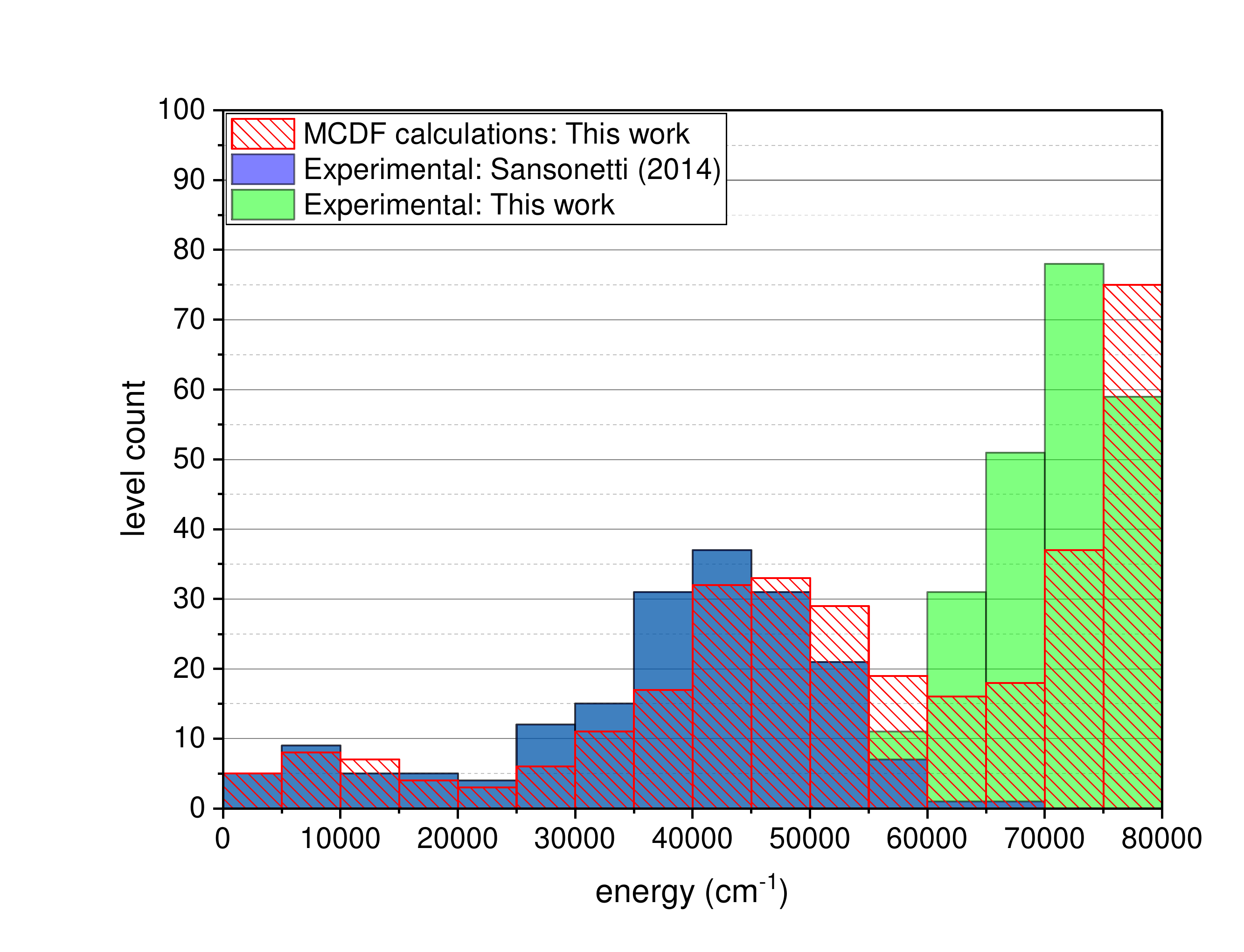}
\caption{Histogram of the number of electronic levels in \Th{}{+} with \J{1/2} to $7/2$ and even parity in the energy range up to 80000 $\mathrm{cm}^{-1}$. The calculated level numbers are shown in red, the measured level numbers in blue (Ref.~\cite{Sansonetti2014}) and green (this work).}
\label{fig:levelstatall}
\end{figure}

With this setup we achieve a good convergence of the calculated level density, however, it has to be noted that the accuracy of our level energies does not go beyond the chosen \lvl{5000}{} energy interval. In Fig.~\ref{fig:levelstatall} we show a comparison of the number of known electronic levels from the database of Ref.~\cite{Sansonetti2014}, our investigations (including Ref.~\cite{Herrera2013}) and the MCDF calculations for even parity levels with \J{1/2} to $7/2$. The calculated level density shows a qualitative agreement with the measurements apart from a prevailing shift to higher energies. The most apparent feature of the level density plot is the minimum around \lvlx{60000}{}. The existence of this minimum comes with an experimental uncertainty, because it lies on the border of the covered energy range of both measurements. Our calculations, however, reproduce this qualitative behaviour, confirming the experimental findings.

Based on our calculations and the experimental findings we can estimate the enhancement factor $\beta$ for the excitation of the \Th{229}{} nuclear transition via an electronic bridge process~\cite{Porsev2010}. Following Ref.~\cite{Dzuba2010} we find the average energy interval between two levels in the desired energy range between \lvl{60000}{} and \lvl{65000}{} to be $160~\mathrm{cm}^{-1} < D_p < 280~\mathrm{cm}^{-1}$. Assuming that the electronic bridge starts from the $J_i=3/2$ ground state of \Th{}{+} via an even parity intermediate level with angular momentum in the range $1/2 < J_n < 7/2$, and with the present value and uncertainty for the nuclear transition frequency $\omega_N$,  we get: $0.5 \times 10^3 < \beta < 6 \times 10^3$.
However this value strongly depends on $\omega_N$ and increases rapidly for $\omega_N > 70000~\mathrm{cm}^{-1}$. A similar result has recently been obtained for the electronic bridge decay rate in Th$^+$ \cite{Karpeshin2018}.

\section{Conclusion}

We have measured the energies of 166 previously unknown high-lying levels in the energy range from \SI{7.8}{} to \SI{9.8}{eV} in \Th{232}{+} and have determined the angular momenta of many of the newly found levels, as well as for previously unclassified levels reported in Ref.~\cite{Herrera2013}. We furthermore have performed MCDF calculations which allow us to predict the electronic level density with acceptable accuracy in the investigated energy range.

The good agreement between measured and calculated level density gives rise to the assumption that most of the even-parity levels with \J{1/2} to $7/2$ in the energy range from \SI{7.8}{} to \SI{9.8}{eV} were found.

The resonant excitation of the nuclear isomer through high-lying levels in \Th{229}{+} is expected to be enhanced compared to the direct optical excitation~\cite{Tkalya1996, Karpeshin1996, Karpeshin1999, Porsev2010}.
In comparison to the investigated energy range in the previous experiment~\cite{Herrera2013} and ab-initio calculations~\cite{Dzuba2010, Porsev2010a}, the level density measured in this work is significantly higher. Therefore we can expect an increased excitation probability for the nuclear isomer via these newly found high-lying levels.

\begin{acknowledgments}
We thank O. A. Herrera-Sancho and K. Zimmermann for contributions to the experimental setup and S. Klein for assistance in the experiment. We furthermore thank T. Leder, M. Menzel, B. Lipphardt and A. Hoppmann for providing expert technical support. We acknowledge financial support from the European Union's Horizon 2020 Research and Innovation Programme under Grant Agreement No. 664732 (nuClock) and from DFG through CRC 1227 (DQ-mat, project B04).
\end{acknowledgments}


\newpage
\appendix
\section{Excitation pathways and angular momentum determination of high-lying levels}\label{AppendixA}

\begingroup
\LTcapwidth=\linewidth
\begin{longtable}{|cc|ccccc|cccc|cccccc|c|}
\caption{Newly observed high-lying states excited through the intermediate levels listed in Tab.~\ref{tab:config} together with levels observed in our previous experiment~\cite{Herrera2013} (marked with "\textbf{*}"). The symbol "\textbf{x}" marks levels which are successfully excited via the specified intermediate level and the symbol "\textbf{--}" denotes that no excitation is observed. If no symbol is given, the excitation has not been checked. Where possible, the levels are given with their total angular momenta derived from the experiment. The energy of the level marked with ** from Ref.~\cite{Herrera2013} is updated. Levels with angular momentum \J{3/2} are classified unambiguously.}
\label{tab:newlevels} \\
\hline
      &       & \multicolumn{15}{c|}{\textbf{Intermediate levels}}                                                                    &  \\
\hline
\textbf{Level} &       & \multicolumn{5}{c|}{\textbf{J = 1/2}} & \multicolumn{4}{c|}{\textbf{J = 3/2}} & \multicolumn{6}{c|}{\textbf{J = 5/2}}         &  \\
\boldmath{}\textbf{[cm$^{-1}$]}\unboldmath{} &       & \textbf{a} & \textbf{b} & \textbf{c} & \textbf{d} & \textbf{e} & \textbf{f} & \textbf{g} & \textbf{h} & \textbf{i} & \textbf{j} & \textbf{k} & \textbf{l} & \textbf{m} & \textbf{n} & \textbf{o} & \boldmath{}\textbf{$J_{exp}$}\unboldmath{} \\
\hline
\endfirsthead
\hline
      &       & \multicolumn{15}{c|}{\textbf{Intermediate levels}}                                                                    &  \\
\hline
\textbf{Level} &       & \multicolumn{5}{c|}{\textbf{J = 1/2}} & \multicolumn{4}{c|}{\textbf{J = 3/2}} & \multicolumn{6}{c|}{\textbf{J = 5/2}}         &  \\
\boldmath{}\textbf{[cm$^{-1}$]}\unboldmath{} &       & \textbf{a} & \textbf{b} & \textbf{c} & \textbf{d} & \textbf{e} & \textbf{f} & \textbf{g} & \textbf{h} & \textbf{i} & \textbf{j} & \textbf{k} & \textbf{l} & \textbf{m} & \textbf{n} & \textbf{o} & \boldmath{}\textbf{$J_{exp}$}\unboldmath{} \\
\hline
\endhead
\hline
\endfoot
\hline \hline
\endlastfoot
	58875.5 & *     & -     &       &       &       &       &       &       &       &       & x     &       &       &       &       &       & 3/2 - 7/2 \bigstrut[t]\\
    59387.1 & *     & -     &       &       &       &       &       &       &       &       & x     &       &       &       &       &       & 3/2 - 7/2 \\
    59477.4 & *     & x     &       &       &       &       &       &       &       &       & x     &       &       &       &       &       & 3/2 \\
    59803.0 & *     & -     &       &       &       &       &       &       &       &       & x     &       &       &       &       &       & 3/2 - 7/2 \\
    60380.1 & *     & x     &       &       &       &       &       &       &       &       & x     &       &       &       &       &       & 3/2 \\
    60618.6 & *     & x     &       &       &       &       &       &       &       &       & x     &       &       &       &       &       & 3/2 \\
    60721.3 & *     & -     &       &       &       &       &       &       &       &       & x     &       &       &       &       &       & 3/2 - 7/2 \\
    61032.4 & *     & x     &       &       &       &       & x     &       &       &       & x     &       &       &       &       &       & 3/2 \\
\hline
\end{longtable}
\endgroup

\newpage
\section{List of newly observed lines}\label{AppendixB}

\begingroup
\LTcapwidth=\linewidth
\begin{longtable*}{|ccrl|ccrl|ccrl|}
\caption{Vacuum wavelengths $\lambda$ and their uncertainties $\sigma$ of the newly found lines in \Th{232}{+} between electronic levels in the energy range from \SI{7.8}{} to \SI{9.8}{eV}. The classification shows the initial and final level with energy in \cmx.}\\
\hline
    $\lambda$ [nm] & $\sigma$ [pm] & \multicolumn{2}{c|}{Classification} & $\lambda$ [nm] & $\sigma$ [pm] & \multicolumn{2}{c|}{Classification} & $\lambda$ [nm] & $\sigma$ [pm] & \multicolumn{2}{c|}{Classification} \bigstrut\\
\hline
\endfirsthead
\caption{Cont.: Newly found lines in \Th{232}{+}.} \\
\hline
    $\lambda$ [nm] & $\sigma$ [pm] & \multicolumn{2}{c|}{Classification} & $\lambda$ [nm] & $\sigma$ [pm] & \multicolumn{2}{c|}{Classification} & $\lambda$ [nm] & $\sigma$ [pm] & \multicolumn{2}{c|}{Classification} \bigstrut\\
\hline
\endhead
\hline
\endfoot
\hline
\hline
\endlastfoot
	237.3774 & 0.5   & 25027 & --~67154 & 252.6039 & 0.6   & 36582 & --~76169 & 269.2348 & 0.6   & 26965 & --~64108 \bigstrut[t]\\
    237.3854 & 0.6   & 26626 & --~68752 & 252.6795 & 0.5   & 36391 & --~75966 & 269.4283 & 0.6   & 36391 & --~73506 \\
    238.4730 & 0.5   & 26965 & --~68899 & 252.7839 & 0.6   & 28244 & --~67803 & 269.5733 & 0.6   & 36391 & --~73486 \\
    238.5148 & 0.5   & 37130 & --~79056 & 252.9017 & 0.6   & 36582 & --~76123 & 269.7499 & 0.5   & 37130 & --~74202 \\
    238.9628 & 0.6   & 26965 & --~68813 & 253.1719 & 0.5   & 36391 & --~75889 & 269.8782 & 0.6   & 26626 & --~63680 \\
    239.2760 & 0.6   & 28244 & --~70037 & 253.2042 & 0.9   & 28244 & --~67738 & 269.9864 & 0.5   & 27403 & --~64442 \\
    239.3099 & 1.5   & 26965 & --~68752 & 253.3508 & 0.6   & 36688 & --~76159 & 270.0000 & 0.6   & 36391 & --~73428 \\
    239.8737 & 0.5   & 36391 & --~78079 & 253.3928 & 0.9   & 26965 & --~66430 & 270.0321 & 0.8   & 36688 & --~73721 \\
\end{longtable*}%
\endgroup


%

\end{document}